\documentclass[aps,pre,twocolumn,amsmath,amssymb,superscriptaddress,showpacs,floatfix]{revtex4-1}
\usepackage{graphicx}
\usepackage{subfigure}

\begin{document}

\title{Percolation perspective on sites not visited by a random walk in two dimensions}
\author{Amit Federbush}\email{afederbush@gmail.com}
\affiliation{Raymond and Beverly Sackler School of Physics and
Astronomy, Tel Aviv University, Tel Aviv 69978, Israel}
\author{Yacov Kantor}\email{kantor@tauex.tau.ac.il}
\affiliation{Raymond and Beverly Sackler School of Physics and
Astronomy, Tel Aviv University, Tel Aviv 69978, Israel}
\date{\today}

\begin{abstract}
We consider the percolation problem of sites on an $L\times L$ square lattice
with periodic boundary conditions which were {\em unvisited} by a random walk
of  $N=uL^2$ steps, i.e., are {\em vacant}. Most of the results are obtained
from numerical simulations. Unlike its higher-dimensional counterparts,
this problem has no sharp percolation threshold and the spanning (percolation)
probability is a smooth function monotonically decreasing with $u$. The
clusters of vacant sites are {\em not} fractal but have fractal boundaries
of dimension 4/3. The lattice size $L$ is the only large length scale
in this problem. The typical mass (number of sites $s$) in the largest cluster
is proportional to $L^2$, and the mean mass of the remaining (smaller) clusters
is also proportional to $L^2$. The normalized (per site) density $n_s$ of
clusters of size (mass) $s$ is proportional to $s^{-\tau}$, while the volume fraction $P_k$
occupied by the $k$th largest cluster scales as $k^{-q}$. We put forward a heuristic
argument that $\tau=2$ and $q=1$. However, the numerically measured values
are $\tau\approx1.83$ and $q\approx1.20$. We suggest that these are effective
exponents that drift towards their asymptotic values with increasing $L$ as
slowly as $1/\ln L$ approaches zero.
\end{abstract}

\maketitle

\section{\label{sec:intro}Introduction}

Percolation theory~\cite{Stauffer91,Gremmett99,Saberi15} provides a statistical description
of long-range connectivity in lattices or networks when some of their sites or links
have been removed. First emerging in the context of polymer
sciences~\cite{Flory1941,Stockmayer1944} and spread of a fluid through a porous
medium~\cite{Broadbent1957}, this theory remains a very active field of research
with very diverse applications, ranging from topography~\cite{Saberi2013},
epidemiology~\cite{Grassberger1983,Matamalas2018}, gelation and
colloid science~\cite{Anekal2006,Coniglio1979,Tsurusawa2019,Adam1990},
environmental~\cite{Taubert2018} and urban~\cite{Makse1995} studies, through
more abstract  networks~\cite{Havlin2015,Kalisky2006,Derenyi2005,Callaway2000}
and more~\cite{Saberi15}. In this paper we consider a two-dimensional case of a variant of
the percolation problem, where an initially full lattice has its sites removed by a
single meandering random walk (RW). The three-dimensional version of the problem
models a degradation of a gel by a single  enzyme, or very few enzymes, that break
the crosslinks they  encounter~\cite{Berry00,Fadda03}.

A percolating system can be characterized by the level of its occupation, such
as the fraction  $p$ of sites present on the lattice, or the occupied volume fraction
in a continuous system. (In this paper we consider {\em site} percolation on
hypercubic lattices, but the results equally well apply to lattice bonds or
mixed site-bond problems and other types of lattices.) In the case of lattice
percolation the geometry can be viewed as a collection of {\it clusters}
formed by neighboring (``connected") occupied sites. A cluster is {\em spanning}
if it forms a continuous path between opposing boundaries in a specific
direction. For an infinite system,  the emergence of a spanning cluster can
be characterized as a phase transition: there exists a sharp percolation
threshold $p_c$, such that for $p>p_c$ there exists an infinite spanning
cluster. Both above and below $p_c$ the mean spatial extent (linear size) of
{\em finite} clusters is called correlation length $\xi$. It diverges near
the threshold as $\xi\sim|p-p_c|^{-\nu}$, where the {\em universal} exponent
$\nu$ is independent of microscopic details of the model, but does depend on
the dimensionality $d$ of the system and, possibly, the presence of long-range
correlations. The universality of the critical exponents allows application of
the results of simple models to more realistic and complicated cases.

One of the simpler percolation models is Bernoulli site percolation on a $d$-dimensional
lattice where each lattice site is {\em independently} occupied with probability
$p$. The exponent $\nu$ of the Bernoulli problem decreases from $\nu_B=1$ at the
{\em lower critical dimension} $d=1$ to $\nu_B=1/2$ for $d\ge d_c=6$, i.e., at
and above the {\em upper critical dimension} $d_c$~\cite{Toulouse74}. The generalized
Harris criterion~\cite{Harris74} has been used to show~\cite{Weinrib84} that
percolation models with short-range correlations or with power-law correlations
$\sim 1/r^b$ with large power $b$ also belong to the Bernoulli percolation
universality class. However, if $b<2/\nu_{\rm B}$, then the correlations are
relevant, and $\nu=2/b$~\cite{Weinrib84}. There is a variety of studies of
correlated percolation~\cite{Coniglio09,Gori17,Riordan2011,DSouza15,K_PRB33}.

In space dimension $d$ we can consider an initially full hypercubic lattice
of linear size $L$ (in lattice constants) and number of sites $M=L^d$ the
sites of which are being removed by an $N$-step RW that started at a random position.
Periodic boundary conditions are imposed, i.e., the walker exiting through one
boundary of the lattice reemerges on the opposite boundary. The number of
steps $N$ of the walker is proportional to the volume of the system, i.e.,
$N=uL^d$, with parameter $u$ controlling the length of the walk. In the
case of gel of crosslinked polymers the random walker represents an enzyme
that breaks the crosslinks of a gel that it encounters~\cite{Berry00,Fadda03}.
The object of the study are the {\em vacant} sites {\em not visited} by
the random walker, that represent the surviving crosslinks. The variable $u$ controls
the concentration of vacant sites and naturally replaces $p$ used in the
regular percolation. For $3\le d\le 6$ infinite clusters of vacant sites
appear for $u$ below similar threshold values $u_c\approx 3$~\cite{KK_PRE100}.
Banavar {\em et al.} studied the geometry of the clusters created by the vacant
sites in $d=2$ and $3$~\cite{Banavar85}, while Abete {\em et al.} considered
the critical behavior near the percolation threshold in $d=3$~\cite{Abete04}.
More recently Kantor and Kardar studied the percolation properties of the
problem for $2\le d\le 6$~\cite{KK_PRE100}.

Sites visited by an $N$-step RW on an {\em infinite} lattice are strongly
correlated. The final position of such a walk is at a distance
$r\approx aN^{1/2}$ away from its starting point, where $a$ is the lattice
constant. This means that number of steps (``mass") scales as $N\sim r^2$, and the
fractal dimension \cite{Mandelbrot82} of a RW is $d_f=2$ independently
of the embedding dimension $d$. Therefore, we may expect our problem to
behave differently in $d=2$ than at higher $d$.  A RW can traverse a
{\em finite} lattice of linear size $L$ in $\sim L^2$ steps, and
therefore a walk of $uL^d$ steps traverses (``crosses") the lattice
\begin{equation}
{\cal N}_{\rm cr}\approx uL^{d-2}
\label{eq:Ncr}
\end{equation}
times. For $d\ge3$ the increase in lattice size $L$ increases
${\cal N}_{\rm cr}$, while a strand of RW on every single ``crossing"
leaves sparser ``footprints" on the lattice. (The total density of
visited or vacant sites in $d\ge3$ remains {\em independent} of $L$ and
depends only on $u$.) This makes the ``thermodynamic limits" somewhat
peculiar even for $d\ge3$ since with the increase of $L$, the structure of the
system changes, rather than having more similar pieces being added to it.

On an {\em infinite} lattice the density of sites visited by an $N$-step RW
(for $d\ge3$) within the distance visited by the walk is $N/r^d\sim1/r^{d-2}$.
On a {\em finite} lattice, the sites belonging to different strands of RW created
due to periodicity of the lattice are almost uncorrelated. The repeated
crossings only contribute to the uncorrelated density of sites. However, the
correlation is preserved for $r$ smaller than the lattice size for sites
situated on the same strand of the RW. Consequently, the {\em cumulant} of the
correlation (from which the overall background density has been subtracted)
has the same power-law relation. Consider a random variable $v(\vec{x})$ which
is 1 if the site at position $\vec{x}$ is {\em vacant}, and zero otherwise.
It is complementary to the variable representing the {\em visited} site
(their sum is 1) and therefore, it has the same cumulant:
$\langle v(\vec{x})v(\vec{y})\rangle_c\sim 1/|\vec{x}-\vec{y}|^{d-2}$~\cite{KK_PRE100}.
Thus, for correlation power $b=d-2$, the correlation length exponent
$\nu=2/b$~\cite{Weinrib84} becomes
\begin{equation}\label{eq:nu}
\nu=2/(d-2),\ \ {\rm for}\ \ 3\le d\le 6.
\end{equation}

The ubiquitous factor ``$d-2$" appearing in the above discussions,
such as in Eqs.~\eqref{eq:Ncr} and \eqref{eq:nu}, or expressions for
the correlation functions, indicates that many of the arguments
presented for $d\ge3$ will change in the two-dimensional case. In this
paper we focus on the vacant site properties in $d=2$ from the point
of view of percolation theory. Despite the absence of percolation
threshold, the system has many scaling properties that resemble critical
phenomena. In Sec.~\ref{sec:2D} we point out the unusual features of
$d=2$, and begin a discussion of two-dimensional vacant site (unvisited
by a RW) percolation (2DVSP) at a point where Ref.~\cite{KK_PRE100} left
off. Furthermore, in Sec.~\ref{sec:2D} we explain the main properties
that set 2DVSP apart from vacant site percolation in higher dimensions
and verify the absence of a sharp percolation threshold in  $d=2$. In
Sec.~\ref{sec:basic}
we consider the mean sizes of the largest cluster and other
clusters, and demonstrate the role played by the lattice size $L$ in
the description of the system. Our results show that $L$ is the main
length scale of the problem, which replaces the correlation length
$\xi$ of other percolation problems. In particular, we show that the
spanning cluster volume and the mean volume of finite clusters both
scale as $L^2$. In Sec.~\ref{sec:fractality} we focus on the geometry
of the spanning cluster and demonstrate that the large clusters in
$d=2$ {\em are not} fractal, contrary to the results of the previous
study~\cite{Banavar85}. Nevertheless, we show that the boundaries of those clusters
are fractal with dimension 4/3. In Sec.~\ref{sec:cluster_statistics} we
study in detail the cluster statistics in 2DVSP.
We put forward a heuristic argument describing cluster statistics
and the effective exponents measured numerically are close to
the proposed theoretical values. We summarize and point out directions
of future research in Sec.~\ref{sec:conclusions}.

\section{\label{sec:2D} How special is $d=2$?}

\begin{figure*}[!ht]
     \subfigure[\label{ClusterBW_a}]{%
      \includegraphics[width=5.5 truecm]{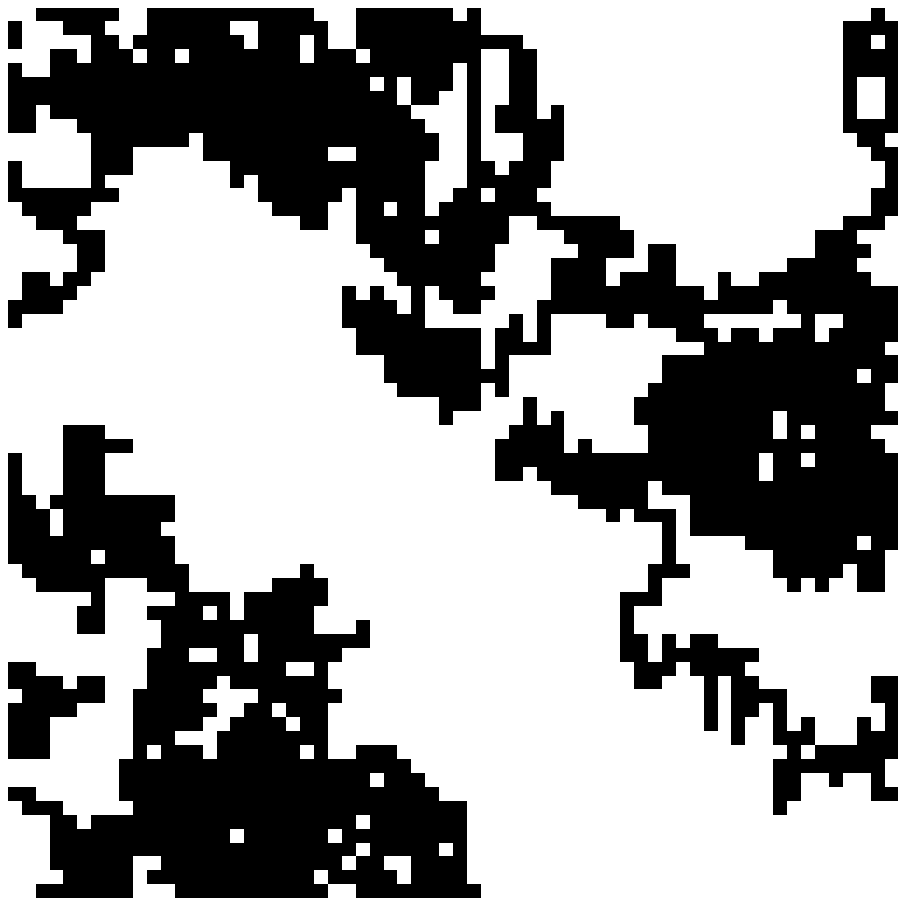}
     }
    \hfill
     \subfigure[\label{ClusterBW_b}]{%
      \includegraphics[width=5.5 truecm]{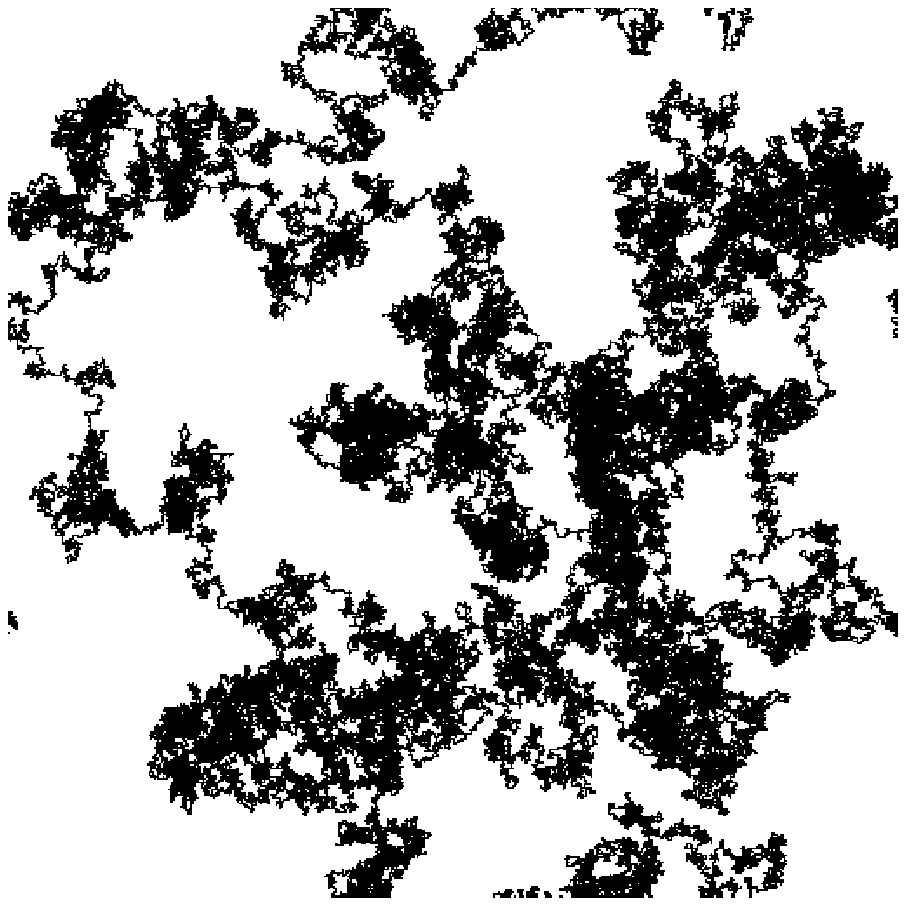}
     }
     \hfill
     \subfigure[\label{ClusterBW_c}]{%
      \includegraphics[width=5.5 truecm]{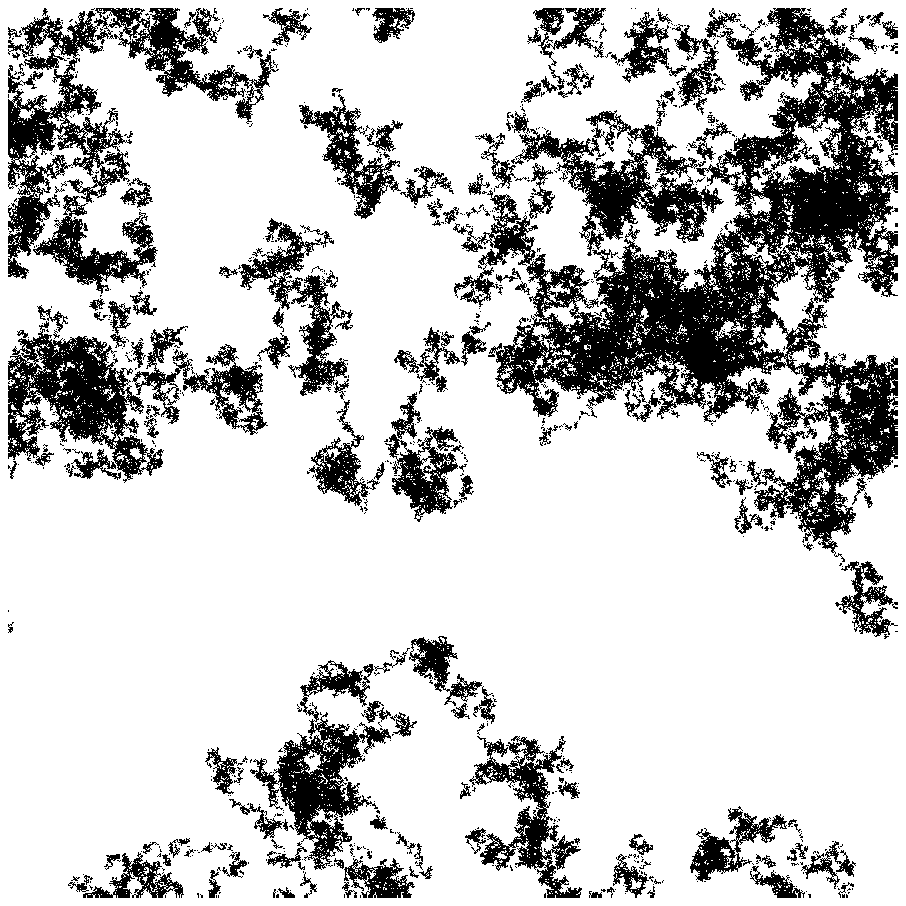}
     }
\caption{Examples of random walks (black) of $N=uL^2$ steps ($u=1.3$) on a
periodic square lattice of size $L^2$ for (a) $L=64$, (b) 512, and (c) 4096.
Despite having {\em the same} RW length parameter $u$, the actual fraction
of the vacant (unvisited) white sites increases with $L$ from (a) $p=0.60$
to (b) 0.70 and to (c) 0.78.
}
     \label{fig:VaryingLdemo}
\end{figure*}

Since the fractal dimension of RW is 2, it ``almost" fills the two-dimensional
embedding space: in $d=2$ the number of {\em distinct} sites visited by an
$N$-step RW  on an {\em infinite} square lattice increases for long walks as
$N_{\rm dist}=\pi N/\ln{N}$~\cite{Dvoretzky51}, i.e., slightly slower than $N$,
because the walk is {\em recurrent} and keeps revisiting previously visited
sites an ever increasing number of times. (In contrast, at $d\ge3$ the number of
distinct visited sites is asymptotically proportional to
$N$~\cite{Vineyard1963,Rubin82} since the number of repeated visits approaches a
constant \cite{Polya1921,Hughes_bookV1}.) The presence of the logarithmic correction in
$N_{\rm dist}$ has consequences for a RW of $N=uL^2$ steps on a {\em finite} square
lattice of linear size $L$ ({$0\leq${$x_{1},x_{2}$}$\leq L-1$}) and volume $M=L^2$.
We assume periodic boundary conditions in {\em both} directions, i.e., the
coordinate $x_i=L$ coincides with $x_i=0$ for $i=1,2$. On such a lattice, it has
been proven \cite{Brummelhuis92} (see also Ref.~\cite{KK_PRE100}) that the mean fraction of
unvisited (vacant) sites for large $L$ is
 \begin{equation}
    p=\exp{\left(-\frac{\pi u}{2\ln{L}}\right)}\,.
    \label{eq:p_u}
 \end{equation}
(For a finite square lattice of $M$ sites, a RW needs on the average
$\frac{1}{\pi}M\ln^2M$ steps to visit {\em all} the sites of the
lattice~\cite{Aldous83,Nemirovsky90,Brummelhuis91,Dembo04,Mendonca11,Grassberger17b}.
The area of complete or nearly complete coverage~\cite{Brummelhuis92,Dembo06a,Comets13} of a
lattice, exhibits unusual effects of {\em discreetness}, such as correlations between
unvisited points, and has been extensively studied. If one just concentrates
on the density of unvisited sites and disregards their connectivity or clusters,
it is possible to identify a fractal behavior with a power-law that depends on
the extent to coverage. However, this limit corresponds to
$u=\infty$ in our problem, while we are concerned only with finite $u$s.)

\begin{figure}[b]
\includegraphics[width=8 truecm]{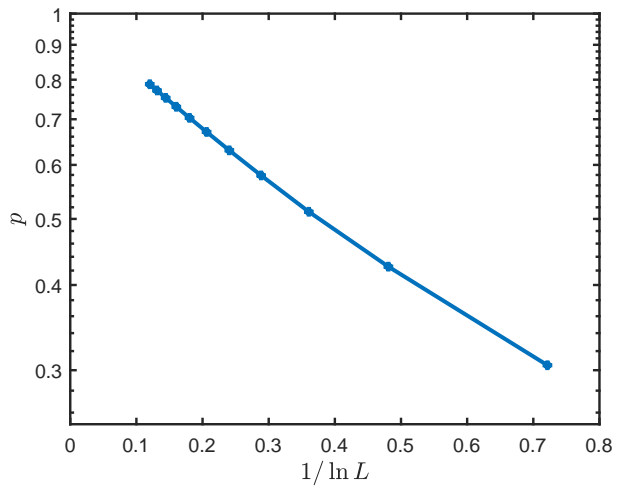}
\caption{Semilogarithmic plot of numerically measured fraction $p$ of vacant
sites for $u=1.3$ as a function of $1/\ln L$, for $L$ ranging between 4 and
4096. (Statistical errors in $p$ calculated from $3\cdot10^5$ configurations
are smaller than the symbol sizes.)
Fraction $p\to1$ as $(1/\ln L)\to0$, as predicted by Eq.~\eqref{eq:p_u}.}
\label{fig:pvsLnL}
\end{figure}

In Eq.~\eqref{eq:p_u} the fraction $p$ of vacant sites depends on $L$ and it
deprives us of a simple correspondence between $p$ and $u$, which is present
in $d\ge3$, where $p=\exp(-A_du)$ with some constant
$A_d$~\cite{Brummelhuis91,KK_PRE100}. However, we note that for fixed $u$ in
$d=2$ in the  $L\to\infty$ limit, we have $p\to1$. Figure~\ref{fig:VaryingLdemo}
depicts RWs with the same $u$ for three different $L$s, demonstrating the
tendency of increasing $p$ with increasing $L$. For {\em very} large $L$s we
can treat the clusters of vacant sites as being separated by ``thin regions"
of RWs. [The results in Fig.~\ref{fig:VaryingLdemo} as well as in many following
figures are presented for $u=1.3$. They do not differ qualitatively from any other
$u=O(1)$. The convenience of showing the data for this particular value of $u$
is explained later in this section and in
Sec.~\ref{sec:fractality}.] Since Eq.~\eqref{eq:p_u}
is valid only asymptotically for large $L$, we measured numerically
the mean fraction of vacant sites $p$ for fixed $u=1.3$ and increasing $L$.
The results are depicted in the semilogarithmic plot in Fig.~\ref{fig:pvsLnL}, and
at such scale Eq.~\eqref{eq:p_u} should be represented by a straight line.
We see that Eq.~\eqref{eq:p_u} is satisfied already for $L\sim 100$.
However, the limit of $p=1$ is approached slowly: for large $L$ the dependence
is $p\approx1-\pi u/2\ln{L}+\dots$. Even for $L=512$ and $u=1.3$, the fraction
of vacant sites $p\approx 0.72$, while for $L=10^6$ we only have $p\approx 0.86$,
and the regime of $p\approx 1$ is numerically inaccessible to us.

When Bernoulli percolation is formulated on, say, a $d$-dimensional hypercubic
lattice, the dimensionless percolation threshold $p_c$ is reached when a sufficient
number of sites is added to an empty lattice or a sufficient number is removed
from a full lattice. In continuum percolation this corresponds to a finite
fraction of the system volume being occupied or removed. When a percolating
situation is created by a RW removing parts of the system, we may separately
consider the length $\ell$ of the single step of the RW and the volume $a^d$
occupied by a certain position $\vec{r}_i$ of the step of the RW. On a lattice
the ``size" of the site $a$ and the length $\ell$ of the step are both assumed
to be equal to the lattice constant. Thus, at any $d$ a {\em short} $N$-step
RW occupies a volume proportional to $Na^d$  on a lattice of volume $(aL)^d$.
If the RW performs $N=uL^d$ steps, then in $d\ge3$ the fraction of vacant sites
is $p=\exp(-A_du)$. This means that for $d\ge3$ there should exist a critical
value $u_c$. This has been proven
theoretically~\cite{Dembo06,Benjamini08,Sznitman08,Sidoravicius09,Sznitman10,Cerny11,Balazs15}
and demonstrated numerically~\cite{KK_PRE100}. (Approaches used in some of these works
also implied the absence of threshold in $d=2$.)

It has been mentioned in Sec.~\ref{sec:intro} that for $d\ge3$ the increase
of $L$ for fixed $u$ causes the increase in lattice crossings by the RW as
expected from the $L$ dependence of ${\cal N}_{\rm cr}$ in Eq.~\eqref{eq:Ncr}.
In $d=2$ the situation is very different:   on one hand, the value of $u$ no
longer solely determines the fraction of occupied or vacant sites due to
$L$ dependence in Eq.~\eqref{eq:p_u}. On the other hand,  the number of
lattice crossings by the RW only depends on $u$ and not on the lattice size
$L$. The probability of having a spanning cluster in, say, vertical direction
depends on the ability of the RW to create a continuous path blocking vertical
connection and is independent of the two-dimensional ``volume"
occupied by the RW. In the absence of a lattice the presence of a ``blocking
path" will depend on the typical step size $\ell$ of the RW rather than
volume (area) $a^2$ occupied by each position.

\begin{figure}[t]
\includegraphics[width=8 truecm]{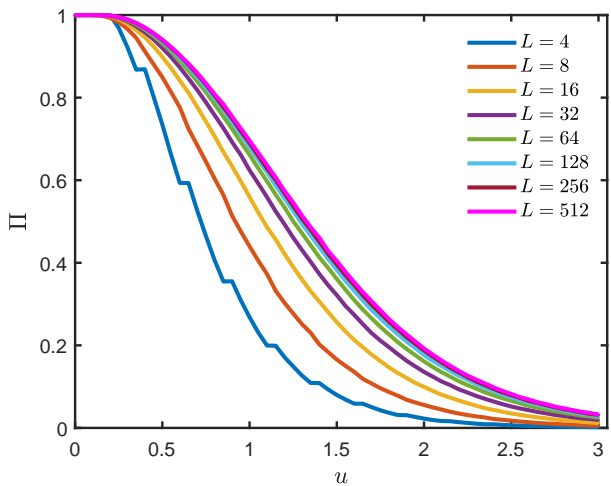}
\caption{Percolation probability $\Pi(L,u)$ for $L=4,8,\dots,512$
(bottom left to top right). Each point is an average of $5\cdot10^{5}$
configurations and the points are separated by $\Delta u=0.05$.}
\label{fig:Pi_by_u}
\end{figure}

Numerically, we consider site percolation on a {\em periodic} square lattice
of $L^{2}$ sites. A random walker starts at an arbitrary site and performs
$N=uL^{2}$ steps with $u=O(1)$. If there is a continuous path of {\em vacant}
sites (unvisited by the RW) that connects the top and bottom boundaries
$x_{2}=0$ and $L-1$, we say that the configuration is {\it spanning}
(percolating). Figure~\ref{fig:Pi_by_u} depicts
the spanning probability $\Pi$ as a function of $u$, for lattice sizes $L$
ranging from $4$ to $512$. (The steps visible on the graph for $L=4$ are a
result of truncating $uL^2=16u$ to an integer.) We see that the graphs of
$\Pi(L,u)$ converge as $L$ increases with the limit being a {\em smooth}
function $\Pi(\infty,u)$, thus indicating that there is no percolation threshold.
(The graphs in Fig.~\ref{fig:Pi_by_u} are similar to the results in Ref.~\cite{KK_PRE100}.)
This differs from systems with a sharp percolation threshold such as vacant
site percolation in $d\geq3$ or Bernoulli percolation, where for $L\gg\xi$,
$\Pi$ rapidly decreases from $1$ to $0$ around the percolation threshold and
becomes a step function in the $L\to\infty$ limit.

The continuous curve in Fig.~\ref{fig:Pi_by_u} demonstrates the absence of a
transition in $d=2$. Due to the central limit theorem, very long lattice
RWs~\cite{rudnick_book} can, on a coarse-grained
level, be treated as Gaussian RWs in continuous space~\cite{Hughes_bookV1}
when the probability of a particular configuration $\{\vec{r}_i\}$ of step positions
of an $N=uL^2$ step walk is proportional to
$\exp\left[-\sum_{i=1}^N(\vec{r}_i-\vec{r}_{i-1})^2/\ell^2\right]$.
(Corrections to Gaussian behavior are irrelevant in the renormalization group sense
and their relative strength decays upon repeated coarse-graining~\cite{Amit93_book}.)
The decrease of $L$ by, say, a factor $\lambda$, and $N$ by a factor $\lambda^2$,
is equivalent to simply integrating out $\lambda^2-1$ out of every $\lambda^2$
variables $\vec{r}_i$ (decimation), which leads to the same functional shape of
the probability with a simple replacement of $\ell$ by $\lambda\ell$
without any change of the system volume. Thus, the Gaussian RW is exactly coarse-grained
without changing the overall distances between points. Not surprisingly, such an
operation usually does not change the spanning cluster, except for slight
coarse-graining of its boundaries.
(The ``exact" nature of coarse-graining transformation applies only to the distances,
but may have nontrivial topological issues, such as breaking apart of a spanning
cluster which was weakly connected before the transformation, or joining of two
nearby clusters and creating a spanning cluster. While such events are unlikely,
they render the entire argument inexact. An example of topological issues of RWs
can be found in the problem
of a winding number - see Ref.~\cite{Drossel96} and references therein.)
The above argument can equally well be used in the opposite direction for
{\em fine-graining} the system by a factor $\lambda$, which involves a replacement
of every bond of a Gaussian RW, by $\lambda^2$ shorter steps with $\ell$
replaced by $\ell/\lambda$. This fine-graining corresponds to an increase of
overall system size from $L$ to $\lambda L$. The $L$ dependence visible in
Fig.~\ref{fig:Pi_by_u} for $L\lesssim 100$ is a consequence of the transition from
a RW on a discrete lattice to an essentially  continuous Gaussian RW  behavior.
Examination of the $L$ dependence of $\Pi$ for a fixed $u$ shows that the
curves in Fig.~\ref{fig:Pi_by_u} almost reached their asymptotic values.

We can interpret the spanning probability $\Pi$ as an average of
a random variable $\Pi'(N)$ which has the value $1$ when there is a spanning cluster
and zero otherwise, when an $N$-step walk has been generated. For a specific
realization of a RW the variable $\Pi'=1$ as the walk begins, and at some step
$N_{0}$ the RW disconnects the spanning cluster, and the system no longer percolates,
i.e., $\Pi'=0$ for $N\ge N_0$. The ``discrete derivative" of this function is
$\frac{\Delta\Pi'}{\Delta N}=-\delta_{NN_{0}}$, where $\delta_{ij}$ is the Kronecker
delta. For large $L$, this can be written in terms of the continuous variable $u$ as
$\frac{d\Pi'}{du}=-\delta(u-u_0)$, where $\delta(x)$ is the Dirac delta-function and
the percolation stops after exactly $u_0L^2$ steps for that configuration.
Clearly, the ensemble averaging over all possible RWs corresponding to a given
$u$ results in the equality   $\left|\frac{d\Pi}{du}\right|=
-\left\langle \frac{d\Pi'}{du} \right\rangle=\left\langle\delta(u-u_0)\right\rangle$,
which is exactly the probability distribution of the percolation stopping times $u_0$.
Calculating $\left|\frac{d\Pi}{du}\right|$ from the data in Fig.~\ref{fig:Pi_by_u},
we find that for large $L$s the distribution of percolation stopping times converges
to a broad peak centered around $u=u^*=1.3$ with half-width of approximately $0.8$.
For every RW, one step before the span breaking
number $N_0$ is reached, the spanning cluster contains at least one ``bottleneck"
that will be pinched off at the next step. Thus, the derivative $d\Pi/du$ can
also be viewed as characterizing such situations.

All the special features of the 2DVSP problem described in this section
qualify $d=2$ to be called the {\em lower critical dimension} of
the vacant site percolation problem. However, it qualitatively differs
from the lower critical dimension of Bernoulli percolation~\cite{Stauffer91}.
For the latter, $d=1$ is the lower critical dimension with a trivial
percolation threshold ($p_c=1$) and various properties that can be
calculated analytically. Besides being a relatively simple problem,
Bernoulli percolation for $p<1$ in $d=1$ has a {\em finite correlation length}
$\xi$ that simply depends on $p$, and the system becomes homogeneous beyond that
length scale. We shall see that 2DVSP does not have such a length scale,
and its structure keeps changing with the increase of the system size $L$.

\section{\label{sec:basic} Mean cluster sizes}

Most quantitative features of percolating systems are extracted from the
shapes and sizes of {\em clusters} of neighboring sites. We identify the
clusters of vacant sites (unvisited by RW) using a Hoshen--Kopelman
algorithm~\cite{Hoshen76}, which efficiently groups the sites into clusters in a
single pass through the lattice. To generate  each configuration we consider a
RW meandering on a lattice of linear size $L$ with periodic boundary conditions
in both $x_1$ (``horizontal") and $x_2$ (``vertical") directions. However, for
the purpose  of cluster identification, we assume that only the $x_1$ coordinate
is periodic, i.e., the clusters can connect through the right and left edges of
the lattice, while the $x_2$ coordinate is not periodic and clusters cannot
connect through the bottom and top edges. The configurations in our simulation
are generated by RWs
of length $uL^2$. For each $u$ and $L$ pair (in a broad range of values) we
simulate a large number of independent realizations, and for each realization
we identify the clusters. We note, that in Bernoulli percolation it can be shown
that usually (in $d\le6$) the infinite cluster is unique at the threshold, but
for finite $L$ we may accidentally have few spanning clusters although the frequency
of such occupancies decreases as a negative power of $L$. In 2DVSP each
configuration is generated by a single continuous RW which tends to create a
single spanning cluster. However, the combination of different boundary conditions
for RWs and for cluster identification, as described above, makes it possible for
finite $L$ to have exceptional (and rare) configurations with more than one
cluster.

\begin{figure}[t]
\includegraphics[width=8 truecm]{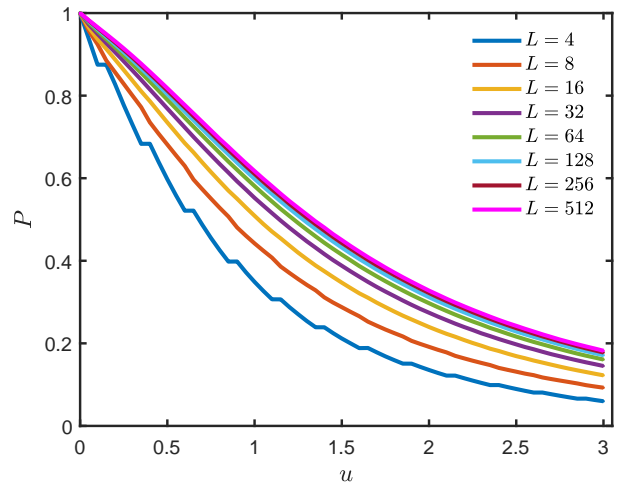}
\caption{Largest cluster strength $P$ dependence on RW length parameter $u$ for
$L=4,8,\dots,512$ (bottom left to top right). Each point is an average of
$5\cdot10^{5}$ configurations and the points are separated by $\Delta u=0.05$.}
\label{fig:P_by_u}
\end{figure}

We denote the probability that a given site belongs to the largest cluster in
the system as the largest {\em cluster strength} $P$. This is an ensemble average
of the number of sites in the largest cluster divided by $L^2$. In Bernoulli
percolation~\cite{Stauffer91} in the $L\to\infty$ limit, below the percolation
threshold $P\to0$ since the largest cluster is finite, while above the
threshold $P$ is finite and represents the volume fraction of the {\em infinite
cluster}. (Therefore, $P$ is used as an order parameter in many percolation
problems.) Moreover, in Bernoulli percolation the concepts of infinite cluster
and spanning cluster coincide. This is {\em not} the case in our problem. Due
to the absence of a percolation threshold, the strength of the largest cluster $P$
includes both spanning and nonspanning clusters. In fact there is no
significant difference in the volumes of both types of largest clusters and
the volumes of all of them are
proportional to $L^2$ leading to a finite $P$. Clearly, $P\leq p$ but this is a
weak bound since $p\to1$ as $L$ increases. Figure~\ref{fig:P_by_u} depicts the largest
cluster strength as a function of $u$ for lattice sizes $L$ ranging from 4 to 512.
(As in Fig.~\ref{fig:Pi_by_u}, the steps for $L=4$ are due to truncation of $uL^{2}$
to an integer value.) As expected, $P(L,0)=1$ since the entire lattice is a single
(largest) cluster, and the function monotonically decreases with increasing $u$. We
observe that the graphs in Fig.~\ref{fig:P_by_u} converge as
$L\to\infty$ to a smooth function $P(\infty,u)$, confirming that the
volume of the largest cluster scales as $L^2$. This observation will play
an important role in Sec. \ref{sec:fractality}. The convergence of the curves
to $P(\infty,u)$ is significantly slower than the convergence of $\Pi$ in
Fig.~\ref{fig:Pi_by_u}, since it is influenced by the slow approach of $p$ to unity,
as indicated by Eq.~\eqref{eq:p_u}: The analysis of the data for a single value of
$u=u^*$ for larger $L$s shows some weak (but linear) dependence of $P$ on
$1/\ln L$ indicating that the asymptotic value is by some 3\% higher than
the value for $L=512$.

The statistics of smaller clusters are of great interest in percolation problems.
We define all the clusters {\em except} the largest cluster as {\em finite clusters}.
We denote by $N_s$ the number of {\em finite} clusters with volume (number of sites)
$s$ in a particular configuration on a lattice, and define the mean
{\em normalized cluster number} $n_s=\langle N_s\rangle/L^2$, where $\langle\rangle$
denotes average over realizations. The exclusion of the largest cluster from the
statistics resembles a similar definition in Bernoulli percolation~\cite{Stauffer91}.
However, in the latter case, it serves as a technical tool to exclude the infinite
cluster above the percolation threshold, and plays a negligible role below the
threshold where the size of the clusters is limited by $\xi$ and for large $L$ there
are many clusters of similar sizes. Thus, in the ``thermodynamic limit" of Bernoulli
percolation the quantity $n_s$ is simply a function describing the prevalence of
finite clusters. In 2DVSP the largest cluster always has its number of sites
proportional to  $L^2$ as do large ``finite" clusters. From the definition of
$n_s$ we see that $sN_s=sn_sL^2$ is the total number of sites belonging to finite clusters of
size $s$ and therefore obtain the identity
 \begin{equation}
    \sum_s sn_s=p-P.
    \label{eq:ns_p_relation}
 \end{equation}
Unlike the case of Bernoulli percolation, the function $n_s$ may depend on $L$ and it
is not evident that a limiting function exists for large $L$. We thoroughly discuss
this function in Sec.~\ref{sec:cluster_statistics}.

\begin{figure}[t]
\includegraphics[width=8 truecm]{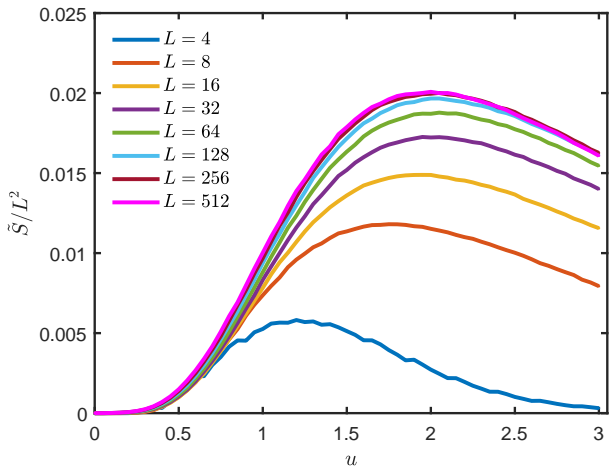}
\caption{Ratio of the mean finite cluster size $\tilde{S}$ and system size
$L^{2}$ as a function of RW length parameter $u$, for $L=4,8,\dots,512$ (bottom to top).
Each point is an average of
$5\cdot10^{5}$ configurations and the points are separated by $\Delta u=0.05$.}
\label{fig:Stilde_by_u}
\end{figure}

The function $n_s$ can be used to determine the mean size of finite clusters: the total
number of lattice sites which belong to an $s$ cluster is $sN_{s}$, and the
total number of vacant sites is $pL^2$, then the probability
that a randomly selected vacant site belongs to a finite $s$ cluster is $sn_s/p$, and the mean
{\em finite} cluster size $\tilde{S}$ is
 \begin{equation}
    \tilde{S}=\frac{\sum_s s^2n_s}{p}\ .
    \label{eq:Stilde_def}
 \end{equation}
This definition of the mean cluster size differs from a similar definition in
Ref.~\cite{Stauffer91} only by the exclusion of the largest cluster in each
configuration in the definition of $n_s$.

\begin{figure*}[!ht]
     \subfigure[\label{Cluster_example_a}]{%
      \includegraphics[width=8 truecm]{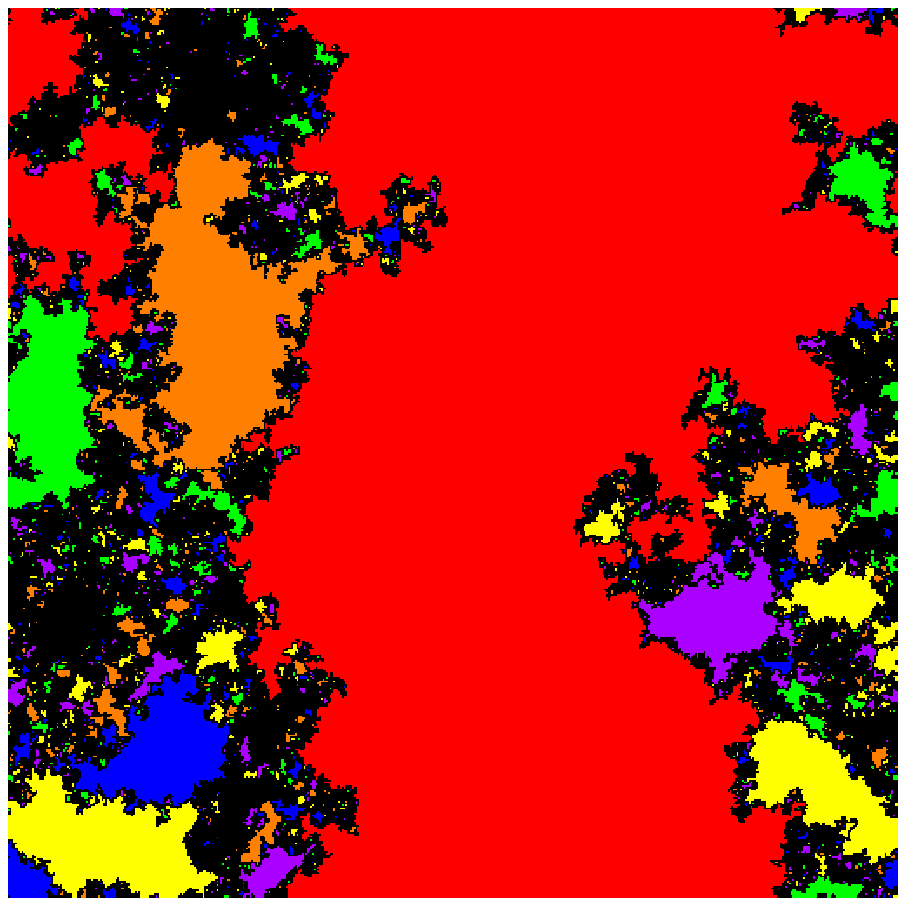}
     }
    \hfill
     \subfigure[\label{Cluster_example_b}]{%
      \includegraphics[width=8 truecm]{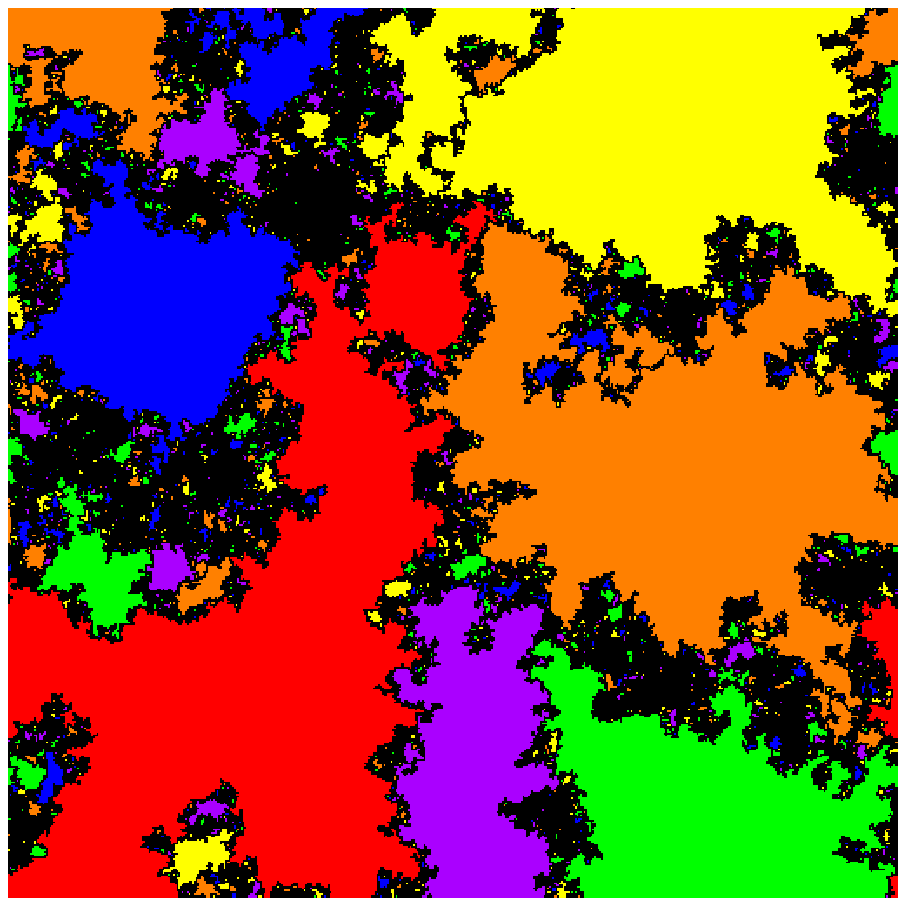}
     }
     \hfill
     \subfigure[\label{Cluster_example_c}]{%
      \includegraphics[width=8 truecm]{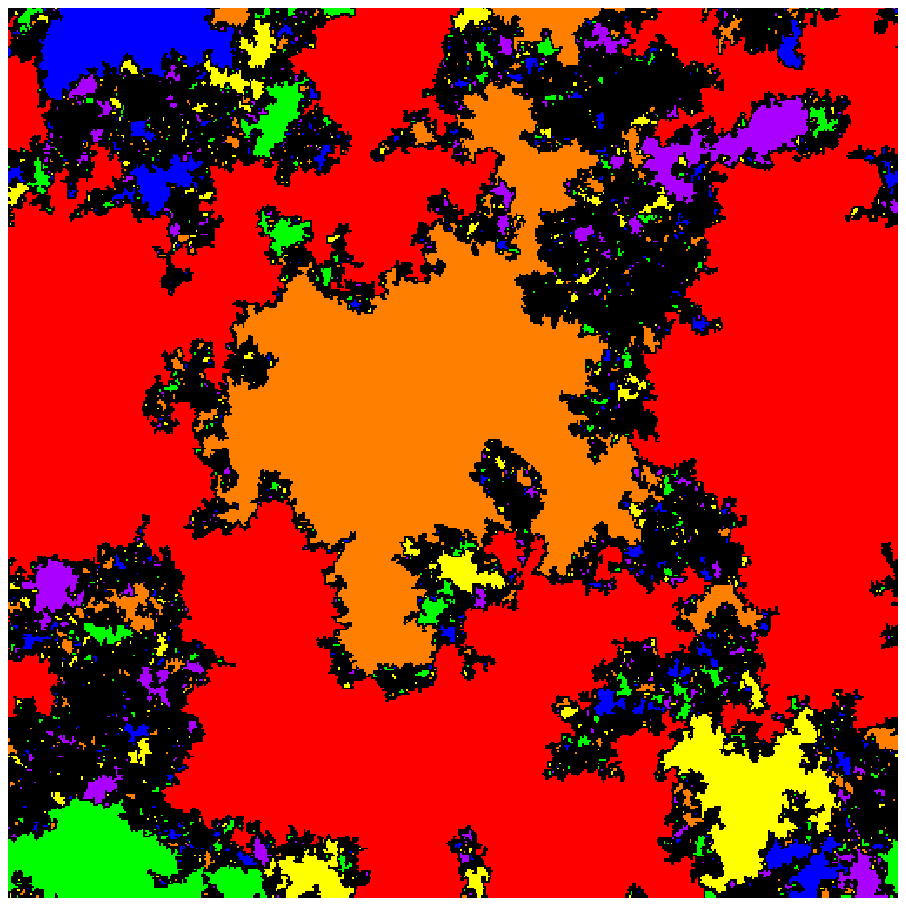}
     }
     \hfill
     \subfigure[\label{Cluster_example_d}]{%
      \includegraphics[width=8 truecm]{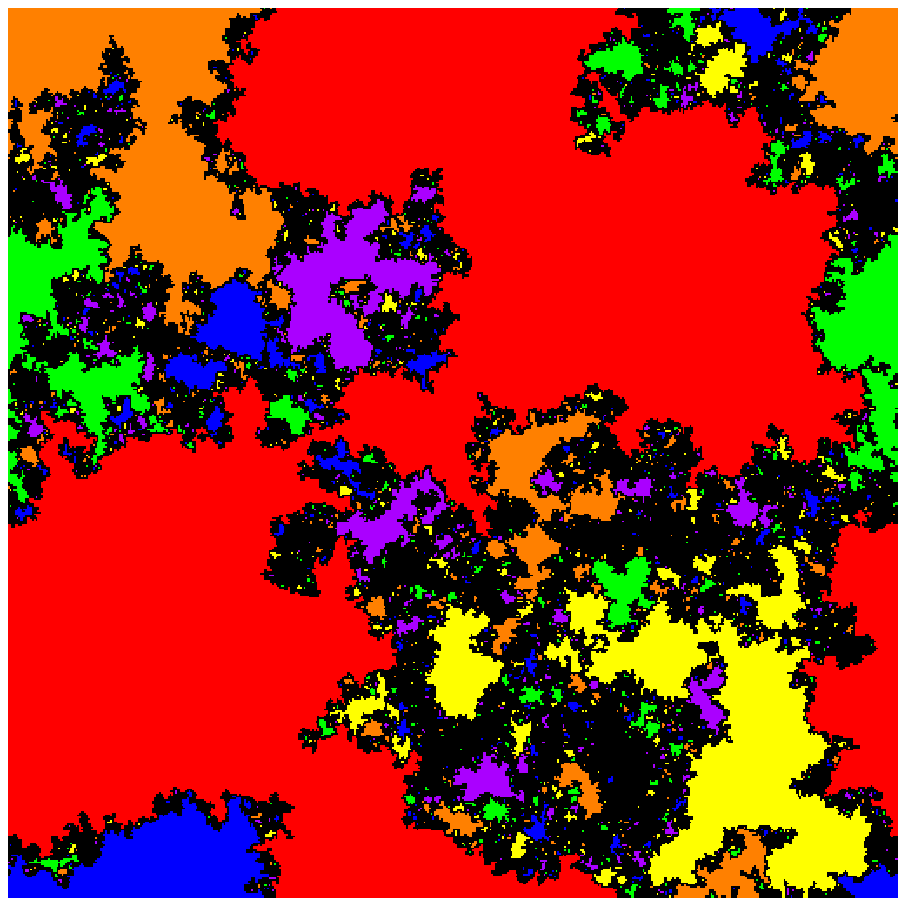}
     }
\caption{Examples of configurations on a $512\times512$ lattice with $u=1.3$. The
sites visited by the random walk are colored black. The clusters of vacant sites
are colored according to their volume: the largest cluster is colored
red, and the rest of the clusters are colored from the second largest
to the smallest according to a periodic color scheme of orange,
yellow, green, blue, violet. (In grayscale format these appears as different shades
of gray with red $\rightarrow$ intermediate gray, blue $\rightarrow$ very dark gray,
violet $\rightarrow$ dark gray, orange $\rightarrow$ light gray, green $\rightarrow$ very light gray,
yellow $\rightarrow$ white.)
The clusters can connect sites through the left and right edges of the lattice
but not through the top and bottom edges. The examples include
(a) a single large percolating cluster,
(b) a nonpercolating system with three large clusters,
(c) a very large and convoluted percolating cluster, and
(d) a large percolating cluster with a narrow bottleneck.
}
     \label{Cluster_example}
\end{figure*}

In regular percolation the mean size of finite clusters is controlled by the
correlation length $\xi$, which increases as the percolation threshold is
approached. However, for $L\gg\xi$ the distribution $n_s$ becomes independent of
$L$. In our problem $\tilde{S}\sim L^2$, and therefore Fig.~\ref{fig:Stilde_by_u}
depicts the {\em ratio} $\tilde{S}/L^{2}$ as a function of $u$ for lattice sizes
$L$ ranging from $4$ to $512$. (As in the previous figures, the steps in the $L=4$
graph are due to truncation to integer $N$.) The curves vanish for $u=0$ since
the entire system is a single largest cluster which is excluded in the calculation
of $n_s$. The values of $\tilde{S}$ increase with increasing $u$ until $u\sim 2$
and then decrease when the RW occupies most of the space for larger $u$. The ratio
$\tilde{S}/L^{2}$ seems to converge as $L\to\infty$, confirming that even the mean
{\em finite} cluster size scales as $L^2$. However, even here the numerical test
of convergence of the function for a single $u=u^*$ indicates that there is a
residual dependence on $1/\ln L$ leading to a slightly larger (up to 3\%) limiting
value of $\tilde{S}/L^2$. From our data it is not possible to determine whether
the maximum of the curves keeps shifting with increasing $L$. Note that the mean
size of the finite clusters at its maximum is only $\sim 0.02L^2$ which is rather
small compared to $P$.

\section{\label{sec:fractality}Geometry and fractality of the largest cluster}

In this section we take a closer look at the geometry of the largest cluster.
Figure~\ref{Cluster_example} depicts four 2DVSP realizations for $L=512$ and
$u=u^{*}=1.3$. This particular value $u=u^{*}$ was selected because it
maximizes $\left|\frac{d\Pi}{du}\right|$, and we expect to see clusters
that are close to the transition between spanning and nonspanning state,
where diverse and ramified configurations can be observed. At $u=u^*$ close
to half of configurations percolate, but the peak in $\left|\frac{d\Pi}{du}\right|$
is very broad and most of the configurations are not very close to the transition
point.

A casual visual inspection of the configurations in Fig.~\ref{Cluster_example}
indicates that the largest clusters have rather ``compact" two-dimensional
interiors and very jagged boundaries. (Similar statements can be made about
the clusters of intermediate sizes.) We also note that the linear dimensions
of large clusters are of the order of $L$. Below we quantify these observations.

In most percolation problems the system is homogeneous beyond the correlation
length $\xi$~\cite{Stauffer91}. However, close to the percolation transition
the correlation length $\xi$, which is the typical linear size of finite
clusters, is much larger than the lattice constant $a$. In the broad range of
distances $a\ll r\ll\xi$, fractal behavior can be observed: e.g., the mass of a
cluster within some distance $r$ from one of its sites increases as $r^{d_f}$,
where $d_f$ is the {\em fractal dimension} of the cluster. Alternatively,
the probability to find a site belonging to the cluster at the distance $r$
from another site of the same cluster decreases as $1/r^{d_{\rm co}}$,
where the fractal {\em codimension} $d_{\rm co}=d-d_f$~\cite{Strelniker09}.
(The relation between $d_f$ and $d_{\rm co}$ is obtained by integrating
the density to find the mass within the radius $r$.)
Thus, the fractal dimension can be measured either by examining total cluster
mass within some distance $r$ or by examining two-point correlation
functions. The presence of fractal behavior is not always easy to ascertain:
E.g., for Bernoulli percolation in $d=2$ the fractal dimension
${d_{f}=\frac{91}{48}\approx1.9}$~\cite{Kapitulnik84,Stauffer91} is not
very different from the embedding dimension.

\begin{figure*}[!ht]
     \subfigure[\label{C_r_a}]{%
      \includegraphics[width=8 truecm]{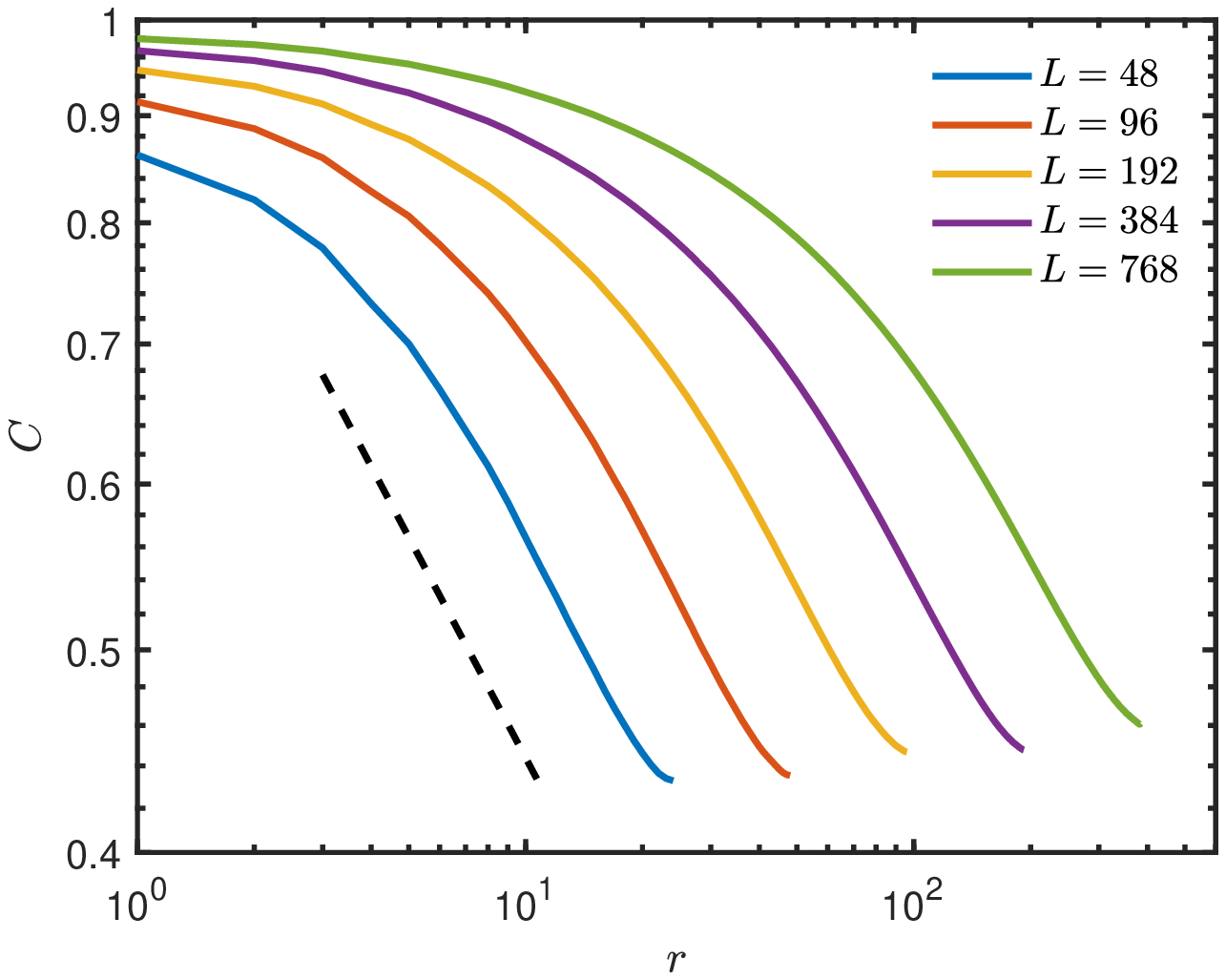}
     }
    \hfill
     \subfigure[\label{C_r_b}]{%
      \includegraphics[width=8 truecm]{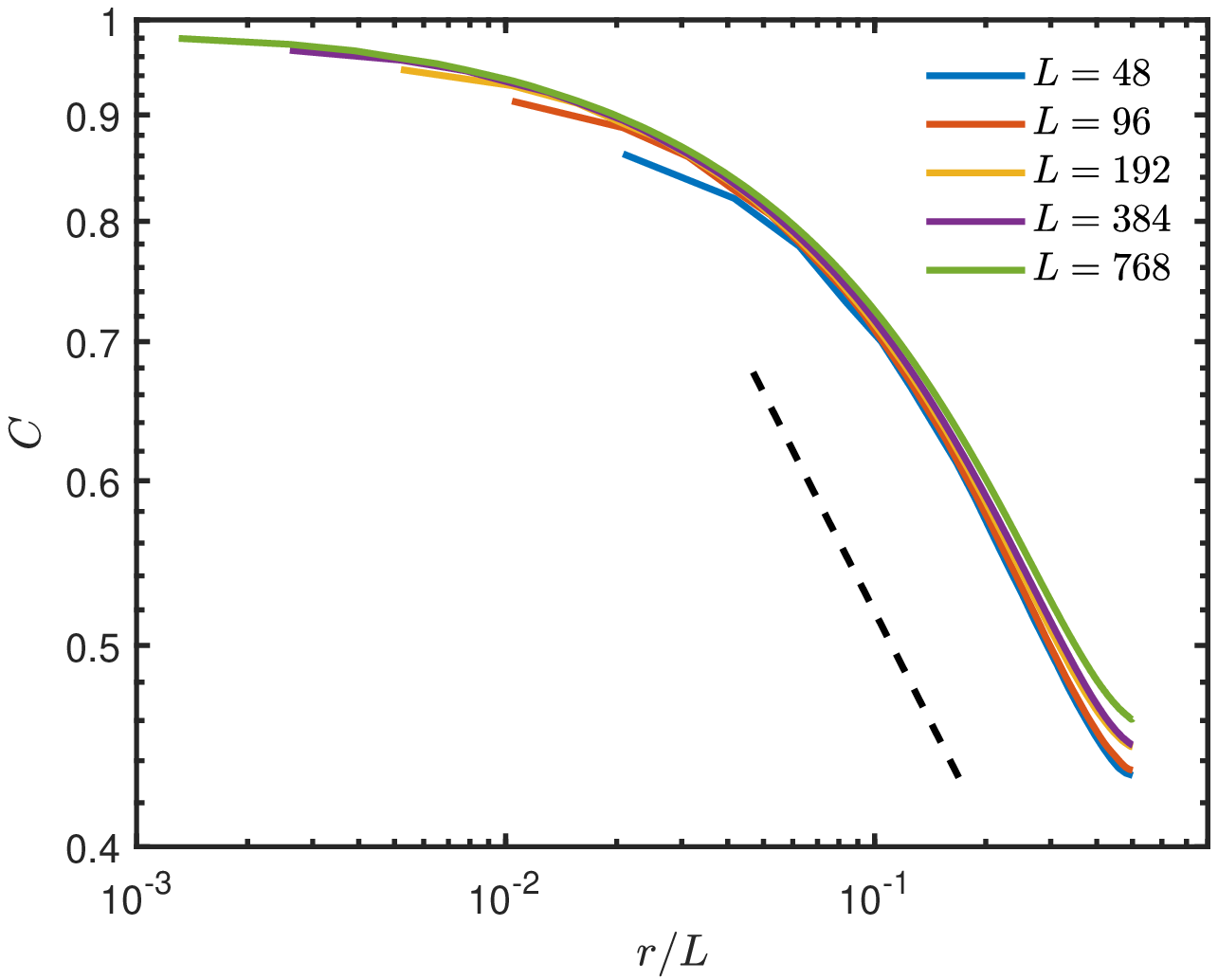}
     }
\caption{Logarithmic plot of the azimuthally averaged two-point correlation $C$
of the largest cluster (see text), averaged over $10^3$ samples, as a function
of (a) distance between the points $r$, or (b) scaled distance $r/L$, for $L$
ranging from $48$ to $768$ [left to right in (a) and bottom to top in (b) with
left-most points of the graphs from left to right].
The dashed line indicates slope $-0.35$.
}
     \label{C_r_graph}
\end{figure*}

In the presence of a percolation threshold, the ``cluster mass {\em versus}
radius" method for measuring the fractal dimension can be reduced to a measurement
(at the threshold) of the mass $PL^d$ of the spanning cluster (part of the
incipient  infinite cluster) as a function of $L$ and equating it to
$L^{d_f}$. Thus, the $L$ dependence  of $P$ contains the information about
$d_f$. In our problem the threshold is absent, while $P$ is independent
of $L$, indicating the absence of fractal behavior, and confirming the
impressions of Fig.~\ref{Cluster_example}. However, in 1985 Banavar
{\it et al.}~\cite{Banavar85} examined two-point correlation functions
of clusters of vacant sites left by a meandering RW at the percolation
point in both $d=2$ and $3$ and reached the conclusion that these
clusters exhibit fractal behavior. In $d=3$ their conclusion should not
be surprising since the system has a percolation threshold and such
behavior is expected. However, in $d=2$ they also found that $d_f=1.75$.
Closely  following their approach (see also \cite{Stauffer91}), we
define the two-point  correlation function for the spanning cluster as
\begin{equation}
C(\vec{r})=\frac{1}{s}\sum_{\vec{r}\mkern2mu\vphantom{r}'}\rho\left(\vec{r}\mkern2mu\vphantom{r}'\right)
    \rho\left(\vec{r}+\vec{r}\mkern2mu\vphantom{r}'\right)\,,
\label{correlation_def}
\end{equation}
where the density $\rho(\vec{r})$ equals $1$ for sites belonging to the spanning cluster
(up to a lattice vector due to periodicity) and zero otherwise, and $\vec{r}\mkern2mu\vphantom{r}'$
is summed over all $s$ sites of the cluster.

Figure~\ref{C_r_graph} depicts the azimuthal average of the correlation function
$C$ for lattice sizes $L$ ranging from $48$ to $768$. All the graphs intersect at
$r=0$ since by definition $C(0)=1$. The data for this figure were simulated using
a different ensemble from the rest of our results: instead of a fixed $u$, the RW
continues until the spanning cluster disconnects, and then we take the configuration
of the lattice before the last step. This method creates an ``almost disconnected"
spanning cluster, and is similar to the methods used in Ref.~\cite{Banavar85}.
(We note that, while the length of the RWs in this ensemble is not fixed, the
typical value of $u$ is $u\sim u^{*}$, consistently with Sec.~\ref{sec:basic}.)
In Fig.~\ref{C_r_a} we see that for smaller fixed $r$, the correlation $C$
approaches  a constant value independent of $r$ as $L$ increases. For larger
$r$, the correlation function $C$ decays exhibiting finite size effects. E.g.,
the value of $r$ at which $C$ drops to, say, $0.6$, doubles every time $L$
is doubled. This is confirmed by the overlapping graphs in Fig.~\ref{C_r_b},
where $C$ is displayed as a function of the scaled variable $r/L$.

Since the absolute value of the slope of each graph in Fig.~\ref{C_r_graph} first
increases for small $r$ and then decreases when $r$ approaches $L$, there is an
intermediate regime on the logarithmic scale (less than $1/3$ of a decade) where
the slope is almost constant, leading to an apparent power-law corresponding
to codimension $d_{\rm co}=0.35$ (indicated by the dashed line). This behavior,
although with a codimension $d_{\rm co}=0.25$ instead, prompted the authors of
Ref.~\cite{Banavar85} to suggest that the spanning cluster is a fractal with dimension
${d_f=1.75}$~\cite{Banavar85}. Their data corresponds to $L=96$, which is the
second left-most graph in Fig.~\ref{C_r_a}. However, in the fractal regime, we
would expect the $C(r)$ graphs corresponding to ever increasing $L$s to be
linear continuations (on the logarithmic scale) of each other with their
cutoffs ever increasing with $L$. Instead we see in
Fig.~\ref{C_r_a} graphs that keep shifting to the right, clearly exhibiting a
finite size ($L$ dependent) effect. We examined this for both the fixed $u$ and
the ``almost disconnected cluster" ensembles, and found no appreciable difference
between the overall behavior of the correlation function. (It should be noted that
when  Ref. \cite{Banavar85} was written, the absence of a critical point of 2DVSP
was not clearly recognized.)
These observations convince us that in $d=2$ the spanning cluster is indeed
{\em compact} and its linear size is proportional to $L$, while its mass is
proportional to $L^2$, consistently with the visual inspection of
Fig.~\ref{Cluster_example}.

\begin{figure}[t]
\includegraphics[width=8 truecm]{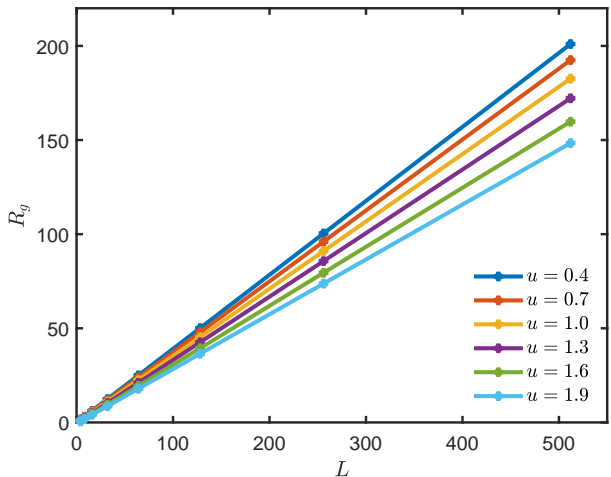}
\caption{Radius of gyration $R_g$ of the largest cluster as a function of
$L$ for $L=4,8,\dots,512$, and for $u=0.4,0.7,\dots,1.9$ (top left to
bottom right). Each point is an average of $3\cdot10^{5}$ configurations.}
\label{fig:Rg_graph}
\end{figure}

The linear size of a cluster can be quantified by its radius of gyration $R_g$, defined
as \cite{Stauffer91,Banavar85}
\begin{equation}
R_g^2=\frac{1}{s}\sum_{i=1}^s(\vec{r}_{i}-\vec{r}_{\rm cm})^2\,,
\label{eq:Rg_def}
\end{equation}
where $s$ is the number of sites in the cluster, $\vec{r}_i$ are the cluster
sites, and $\vec{r}_{\rm cm}$ is the position of the center of mass of the
cluster. It should be noted that due to the periodic boundary conditions in
the horizontal direction both the positions of steps $\{\vec{r}_i\}$
and the position of center of mass $\vec{r}_{\rm cm}$ are not always uniquely
defined. The proper choices are made to minimize the resulting $R_g$.
In regular percolation problems, the mean $R_g$ of the clusters, as well
as $R_g$ of typical large clusters scale as the correlation length $\xi$.

We examined the relation between $R_g$ of various clusters and their mass
for clusters in the entire range of sizes $s$. The particular values of the
linear extent $R_g$ of various clusters with a given specific mass $s$ are
broadly scattered, but the average values of $R_g^2(s)$ are proportional
to $s$, leading to the conclusion that the clusters are {\em not} fractal,
similarly to the largest cluster. Figure~\ref{fig:Rg_graph} depicts $R_g$ of
the largest cluster as a function of $L$, for $L$ ranging from $4$ to
$512$ and for $u=0.4,0.7,...,1.9$. We see that $R_g$ clearly displays
the expected linear scaling with $L$, although the slope of linear curves
in Fig.~\ref{fig:Rg_graph} slowly decreases with increasing $u$.

The jagged boundary of the clusters seen in Fig.~\ref{Cluster_example}
is formed by segments of a RW. We define the cluster hull perimeter $H$
as the total mass of cluster sites bordering the sites visited by the
random walk. We considered the hull perimeter only for the largest cluster.
For fixed $L$ we may expect the numerical value of the perimeter to
decrease as $u\to0$ since the spanning cluster will essentially have
no boundaries, and on the other hand for $u\gg1$ the perimeter will
again be small due to a decrease in the typical size of the largest
cluster. Somewhere at the intermediate values of $u$, possibly
around $u^*$, we will see large hull perimeters.
Unlike Bernoulli percolation, our clusters do {\em not}
have internal boundaries due to holes inside a cluster, and therefore
their entire perimeter belongs to the hull. For Bernoulli site percolation at
$d=2$ and $p=p_c$, both the mass and the hull of the spanning cluster
are fractal, and the hull scales as $H\sim L^{D_H}$ with fractal
dimension $D_{H}=1.74$~\cite{Grossman1986,Voss1984}.

\begin{figure}[t]
\includegraphics[width=8 truecm]{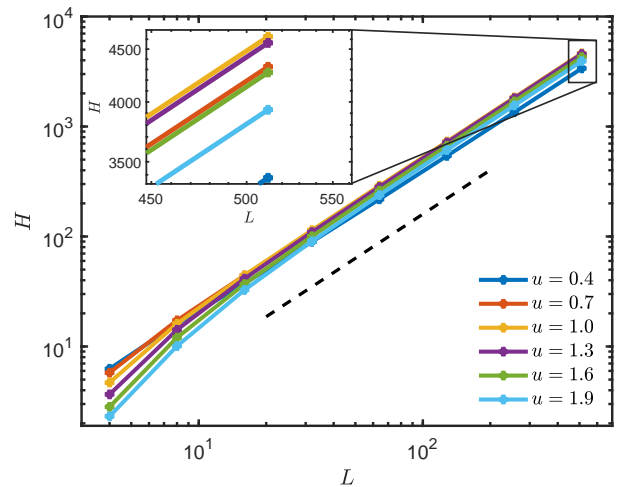}
\caption{Logarithmic plot of the hull perimeter $H$ of the largest cluster
as a function of $L$ for $L=4,8,\dots,512$, and for $u=0.4,0.7,\dots,1.9$
with left-most points of the graphs from top to bottom. Some of the lines intersect,
and the right-most points (shown enlarged in the inset) are ordered from top to bottom
in order $u=1.0,1.3,0.7,1.6,1.9,0.4$, i.e., for larger $L$s the hull perimeter slightly
increases as $u$ increases from 0.4 to 1.0 and then decreases. Each point is an average
of $3\cdot10^{5}$ configurations. The dashed line indicates slope $4/3$.}
\label{fig:Hull_graph}
\end{figure}

Figure~\ref{fig:Hull_graph} depicts the hull perimeter $H$ of the largest
cluster of our problem as a function of $L$, for $L$ ranging from $4$ to
$512$ and for $u=0.4,0.7,...,1.9$ on a logarithmic scale. For a fixed
large $L$ the hull perimeter $H$ slightly depends on $u$ reaching a maximum
close to $u=1.0$ (see inset in the figure). With increasing $L$ all of the graphs approach straight lines
with slopes corresponding to a power $D_H=1.33\pm0.01$, where the size of
the error provides a subjective estimate of uncertainty in extrapolation
as well as slight differences between different values of $u$. The dashed
line in Fig.~\ref{fig:Hull_graph} has a slope of 4/3 and provides a guide to the eye.
(The statistical errors are negligible.) The measured exponent of 4/3 coincides
with the well-known theoretical result proven by Lawler
{\it et al.}~\cite{Lawler2001} (and conjectured by
Mandelbrot~\cite{Mandelbrot82}) that the fractal dimension of the frontier of
a Brownian motion in $d=2$ is $4/3$.

\section{\label{sec:cluster_statistics}Cluster Statistics}

The mean number of clusters of size $s$ per lattice site $n_s$ as defined in
Sec.~\ref{sec:basic} is one of the most revealing features of a percolating system.
Despite the differences between 2DVSP and the usual Bernoulli percolation
we will attempt to follow a similar logic while pointing out important
differences between the systems. If a system lacks a length or mass scale, then we
expect the functions characterizing the system to be power-laws. In
particular, one might expect $n_s=As^{-\tau}$, where $\tau$ is called a
Fisher exponent \cite{Stauffer91}, while $A$ is a {\em constant}, possibly
dependent on some microscopic properties and details. In Bernoulli
percolation such dependence is valid on scales $r$ much larger than
the lattice constant $a$ but smaller than the correlation length $\xi$, i.e.,
for cluster masses satisfying $1\ll s\ll s_c$, where $s_c$ is a typical mass
of a cluster of linear size $\xi$. (Typically, there is
a power-law dependence between $s_c$ and $\xi$.) At length scales $r\sim\xi$
the power-law is corrected by some {\em cutoff function} $F_c$, which is
$\approx 1$ for $s\ll s_c$ and drops to zero as $s_c$ is exceeded. The overall
shape of the dependence is
\begin{equation}
    n_s=A s^{-\tau} F_c \,.
    \label{eq:ns_power}
\end{equation}
It is frequently assumed in Bernoulli percolation that the cutoff function
depends only on the ratio $s/s_c$, although away from the threshold a more
complicated dependence on $p$ might appear~\cite{Ding14}. At the percolation
threshold ($\xi=\infty$) the cutoff is absent, and therefore
Eq.~\eqref{eq:ns_p_relation} dictates that an infinite sum
$\sum_s sn_s\approx\sum_s As^{1-\tau}$ converges, and therefore $\tau>2$.
Indeed for two-dimensional Bernoulli percolation
$\tau=197/81\approx2.05$~\cite{Stauffer91}.

In the 2DVSP problem the function $n_s$ plays a somewhat different role.
In percolation problems with a threshold, there is some correlation length
$\xi$ and therefore a very large system of linear size $L$ can be treated
as a collection of $(L/\xi)^d$ independent systems. Consequently, even a
{\em single} very large realization of the system assures that most cluster
sizes $s$ will be present and $n_s$ can be naturally treated as a continuous
function representing the frequency of clusters of size $s$. In the
2DVSP, there is no correlation length and $L$ is the only large length scale.
In a single sample there are only a few large clusters of size $s\sim L^2$ and,
consequently, if we avoid averaging over samples for most large $s$, we will
have vanishing $n_s$ and only a few particular values of $s$ will produce
$n_s=1/L^2$. An increase of $L$ will not improve that situation.
Only {\em averaging over the ensemble} will produce a continuous function
$n_s$ of $s$. Nevertheless, we  expect to have a large range of scale-free
behavior, and, as in the case of regular percolation, we hope that
the ensemble averaged $n_s$  has a shape given by Eq.~\eqref{eq:ns_power}.

\begin{figure}[t]
\includegraphics[width=8 truecm]{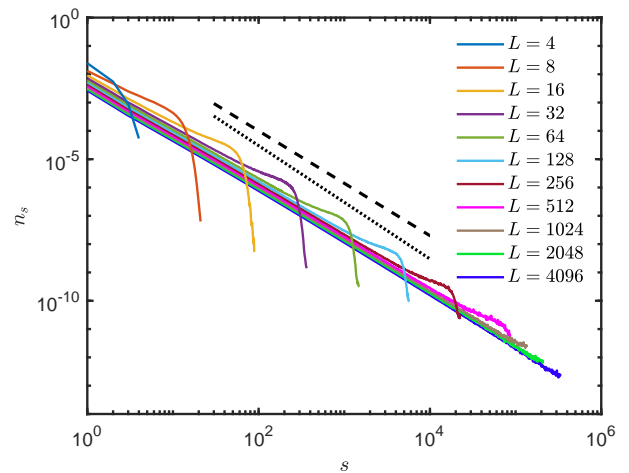}
\caption{Logarithmic plot of $n_s$, averaged over $7\cdot10^5$ samples,
as a function of the cluster mass $s$, for $u=1.3$ and for
$L=4,8,\dots,4096$ (left to right).
[Curves for $L\ge 1024$ are truncated for large $s$ (see text).]
The dashed and dotted lines indicate slopes $-1.85$ and 2, respectively.}
\label{fig:n_s_graph}
\end{figure}

Figure~\ref{fig:n_s_graph} depicts a logarithmic plot of the cluster number
per site $n_s$  for $L$  ranging from $4$ to $4096$ and $u=1.3$. All the
graphs consist of a relatively straight region for $s\ll L^2$ and a cutoff
around $0.3L^2$. Close to the cutoff, the curves exhibit somewhat unusual
behavior described in the next paragraph. We excluded the area close to
the cutoff from our analysis of scaling behavior. While we used
rather large statistical samples, the frequencies of finding a cluster
of some particular (large) $s$ for large $L$s become very low and the
resulting curves are very ``noisy." The curves presented in
Fig.~\ref{fig:n_s_graph} are ``smoothed" by averaging the results for a
particular $s$ over a range $\sim\sqrt{s}$. This procedure has almost
no effect for moderate $L$s, but distorts the ``tails" of the curves
on large lattices: for $L\ge 512$, when $n_s$ drops below a certain (small)
value, where in the {\em entire ensemble} of $7\cdot 10^5$ samples there is
about one cluster of each size $s$, the averaging procedure distorts the
curve, because it is just an averaging  of zeros and ones. We therefore truncate
the curves when this value is reached. For $L=512$ the truncation appears
when the cutoff is reached, while for larger $L$s the graphs are truncated
even before reaching the cutoff.

All the curves for $L<512$ in Fig.~\ref{fig:n_s_graph} exhibit a sharp cutoff
and the position of that  cutoff increases by a factor of 4 every time $L$
doubles. Such $L$ dependence is consistent with the results that we had in
Sec.~\ref{sec:basic}. Therefore, the cutoff function depends of
$s/s_c$, with $s_c\sim L^2$ and with some dependence on $u$. Near the
cutoff position the function $F_c$ has some structure: In
Fig.~\ref{fig:n_s_graph} instead of being just a simple monotonic drop
from 1 to 0, it actually increases above 1 before the drop thus moderating
the decay of $n_s$ close to $s_c$. The behavior close to the cutoff depends
on $u$: For smaller $u$s the ``bump" in $F_c$ becomes even more pronounced.
Such nontrivial shape of $F_c$ apparently reflects the fact that
the large-$s$ part of $n_s$ attempts to depict few very large clusters by
using a smooth function of $s$.

Before the cutoff, the curves in Fig.~\ref{fig:n_s_graph} are fairly straight
and approximately follow a slope the absolute value of which very slowly increases
and reaches the value of $\tau\approx 1.83$ (or slightly larger) for the
largest $L$s.
The dashed line in Fig.~\ref{fig:n_s_graph} indicates a slope of -1.85. Such behavior
is a significant deviation from the expectation that $\tau$ should exceed 2.
If this represents
an asymptotic trend, then the requirement for $\sum_s sn_s$  to be
finite for ever increasing $L$, i.e., with the power-law cutoff increasing as
$L^2$, would require the prefactor of the power-law in Eq.~\eqref{eq:ns_power}
to decrease $A\sim L^{2(\tau-2)}$. While the vertical position of the curves in
Fig.~\ref{fig:n_s_graph} slightly decreases with increasing $L$ it is extremely
weak, possibly dependent on $1/\ln L$. Strong $L$ dependence of $A$ would
also cast doubt on our assumption of scale-independence of the results on the
intermediate scales. The results described in this paragraph, namely
$\tau<2$ and $A$ almost independent of $L$, are not mutually consistent.

\begin{figure}[t]
\includegraphics[width=8 truecm]{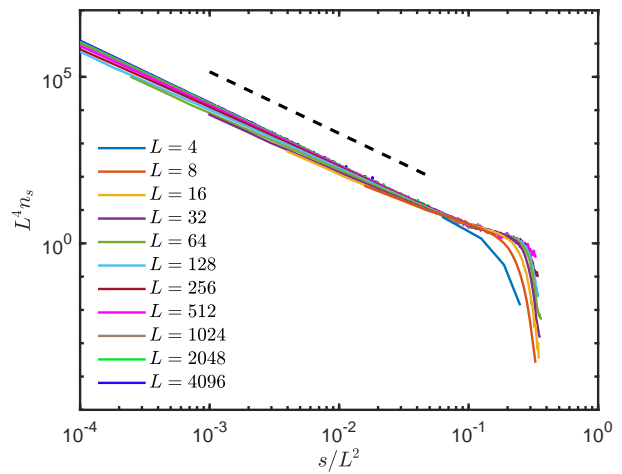}
\caption{Logarithmic plot of $L^{4}n_{s}$, averaged over $7\cdot10^5$ samples, as a function of the scaled cluster mass $s/L^2$, for $u=1.3$ and for $L=4,8,\dots,4096$
(bottom to top, which is left to right at the right ends of the graphs).
[Curves for $L\ge 1024$ are truncated for large $s$ (see text).]
This is a scaled version of the data in Fig.~\ref{fig:n_s_graph}. The dashed line indicates slope $-1.85$.}
\label{fig:ns_scaled}
\end{figure}

Special properties of Gaussian RWs in $d=2$ can be used to advance a
theoretical heuristic argument that the theoretical value of $\tau$
should be $\tau_{\rm th}=2$. Long RWs
on a lattice can be treated as Gaussian RWs in a continuum. As has been
mentioned in Sec.~\ref{sec:2D}, a
Gaussian RW can be coarse-grained or fine-grained exactly:
The increase of lattice size from $L$ to $\lambda L$ and number
of steps $N$ from $uL^2$ to $u\lambda^2L^2$ is equivalent to
keeping the system size unchanged, while increasing $N$ by a factor
$\lambda^2$ and decreasing the step size $\ell$ by factor $\lambda$.
Positions of every $\lambda^2$rd step of this new fine-grained
configuration will be distributed exactly as the positions of the
original Gaussian RW before it was fine-grained. Thus, the process
of fine-graining replaces each step of the RW by $\lambda^2$ smaller
steps but {\em does not change the paths of RWs on larger scales}.
We saw in Sec.~\ref{sec:fractality} that the clusters have compact
interiors, and therefore their volume will not change, except for
being measured in smaller units, i.e., a cluster of $s$ sites will
become a cluster of $\lambda^2s$ sites, of the same shape, although
with more jagged boundaries. This argument, as mentioned in
in Sec.~\ref{sec:2D} assumes that the slight
changes in the fine-grained  boundaries created by the RW do not break
up fragile clusters  that have bottlenecks or join clusters which
were ``almost  connected" in the original geometry, or at least such
changes are very rare and do not modify the distributions.
The change of scale will certainly have effect on creation or elimination
of the smallest clusters. Thus by assuming that nothing
changed in the overall geometry we find that the number of clusters
in a certain range $N_s(L)\Delta s$ remains unchanged in the new units,
i.e., it is equal to $N_{\lambda^2 s}(\lambda L)\Delta(\lambda^2 s)$. This
equality is valid only for large enough $s$, far from the smallest $s$
where the cluster number is dependent on the resolution and graining.
By dividing both terms by $L^2$ and eliminating $\Delta s$ we find that
\begin{equation}
    n_{s}(L)=\lambda^{4}n_{\lambda^2s}(\lambda L).
\label{eq:ns_n_lambda2_s_relation}
\end{equation}
The argument presented above assumed $\lambda>1$ and resulted in fine-graining
of the system. We could use $\lambda<1$ and coarse grain the system.
The resulting relation is valid for arbitrary $\lambda$, and, in particular,
by taking $\lambda=1/L$ we observe that $L^4n_s(L)$ becomes a function
of only of $s/L^2$. Figure~\ref{fig:ns_scaled} depicts the scaled number
of clusters as a function of the scaled cluster mass, and the graphs for various
$L$s almost (but not completely) collapse. Furthermore, the scaled cutoffs are
very similar and appear at about $s_c/L^2\approx 0.3$, yet again confirming
relation $s_c\sim L^2$.

The left hand side of Eq.~\eqref{eq:ns_n_lambda2_s_relation} is independent of
$\lambda$, and the equation must be valid for an arbitrary $\lambda$ in the
power-law regime that is only possible, when $n_s=As^{-2}$ leading to
the ``theoretical" value of the Fisher exponent $\tau_{\rm th}=2$. A slope
of -2 is indicated by the dotted line in Fig.~\ref{fig:n_s_graph}, and seems
to be slightly larger than the approximate slope of -1.83 seen in the graph.
Later we will explore the possibility that our results did not yet converge
to their asymptotic values. If the Fisher exponent is 2, then the coefficient $A$
cannot be constant: to maintain a finite $\sum_ssn_s$, it must decrease as
$1/\ln L$.  Indeed, $A$ slowly decreases with increasing $L$. Our
heuristic argument is not accurate enough to determine the logarithmic
terms either in the prefactor, or even in the $s$ dependence of $n_s$.

\begin{figure}[!t]
\includegraphics[width=8 truecm]{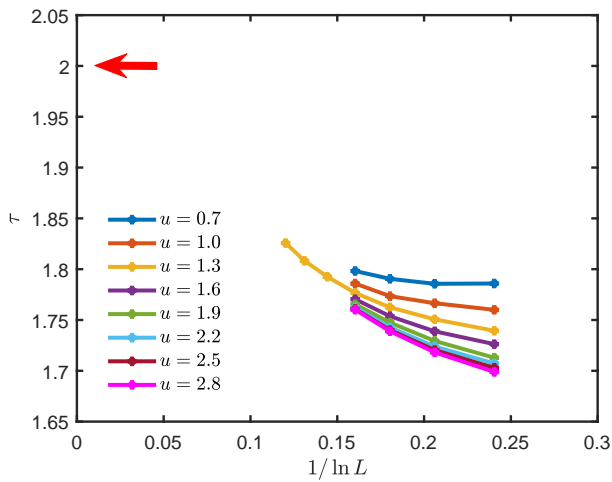}
\caption{Plot of the effective exponents $\tau$  defined in
Eq.~\eqref{eq:ns_power}, as a function of $1/\ln{L}$, for $u=0.7,1.0,...,2.8$
(top to bottom) and  $L=64,128,256,512$. For $u=1.3$ the range of
 $L$s was extended to $L=1024, 2048, 4096$. The exponents
are extracted from linear fits (on logarithmic scale)
in the range $1\le s\le 0.01L^2$ (see text). Estimated systematic errors
in vertical positions of data points are as large as 0.05. The arrow near the
vertical axis indicates the expected theoretical value $\tau_{\rm th}=2$,
rather than a value extrapolated from data.
}
\label{A_tau_fit_graph}
\end{figure}

In Sec.~\ref{sec:basic} we discussed the possibility of extremely slow
convergence of the numerical results when the $L$ dependence may be as slow
as $1/\ln L$. We attempted to study the apparent discrepancy between the
heuristic result $\tau_{\rm th}=2$ and the measured $\tau\approx1.83$.
We  measured the weak dependence on $L$ of the effective exponent $\tau$.
The exponent has been extracted from a linear fit on a logarithmic scale
in the range $1\leq s\leq 0.01L^2$, which avoids the very peculiar behavior
of the curves near the cutoff. The data points for $\tau$ in
Fig.~\ref{A_tau_fit_graph} are arithmetic means of the slopes in the two
halves of the range, $1<s<0.1L$ and $0.1L<s<0.01L^2$.
Figure~\ref{A_tau_fit_graph} depicts $\tau$ as a function of $1/\ln{L}$
for $L=64,128,256,512$ and for ${u=0.7,1.0,\dots,2.8}$. For $u=1.3$ the
simulations have been extended to $L=1024, 2048, 4096$. Slightly nonlinear
behavior (on the logarithmic scale) of the graphs in Fig.~\ref{fig:n_s_graph}
introduces possible systematic errors as large as $0.05$. Such errors,
combined with extremely slow $L$ dependence prevent a reliable extrapolation
to $L\to\infty$. Nevertheless, Fig.~\ref{A_tau_fit_graph} demonstrates
the {\em plausibility} of the asymptotic value $\tau_{\rm th}=2$ that is
indicated by an arrow.

Usage of the continuous (ensemble averaged) function $n_s$ obscures the fact
that there are only a few large clusters, and therefore in each sample there
are no clusters for most large values of $s$. It is therefore beneficial
to examine these statistics from a different point of view. We denote by
$P_k$ the mass (volume) of the $k$th largest cluster, divided by the lattice
volume $L^2$, and averaged over realizations. In particular, the largest cluster
strength $P$ equals $P_1$, and the identity ${\sum_{k}P_{k}=p}$ is trivial.
By construction $P_k$ is monotonous and
should exhibit a more ``continuous" behavior than $n_s$ because for large
clusters (small $k$), $P_k$ should converge in the $L\to\infty$ limit just
as $P$ does, and for smaller clusters the differences between cluster sizes
are small and there are many of them.

\begin{figure}[t]
\includegraphics[width=8 truecm]{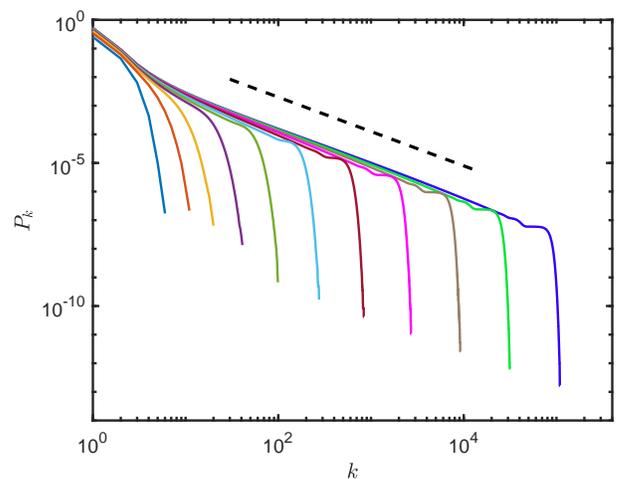}
\caption{Logarithmic plot of $P_{k}$, averaged over $7\cdot10^5$ samples,
as a function of the cluster mass index $k$, for $u=1.3$ and for
$L=4,8,\dots,4096$ (left to right). The dashed line indicates slope $-1.20$.}
\label{fig:Pk_graph}
\end{figure}

Figure~\ref{fig:Pk_graph} depicts $P_k$ as a function of the cluster mass
index $k$, for $L$ ranging from $4$ to $4096$ and $u=1.3$. The graphs are
obtained from the average of $P_k$ over many configurations, in the same
manner as the graphs of $n_s$ in Fig.~\ref{fig:n_s_graph}. For $k>10$ the
graphs appear to roughly converge to a straight line representing a power
law $Bk^{-q}$, up to a sharp cutoff $k_c$ which increases with increasing
$L$. The effective exponent $q$ slightly decreases with increasing $L$ and
reaches $q=1.20$ for the largest sample. (The estimated systematic errors
of the effective exponents are smaller than 0.05.) The prefactor $B$ is
almost independent of $L$. We note
that most of the mass of the vacant clusters is contained in three or four
largest clusters, while the remainder contains a small fraction of vacant
sites. Since large $k$ corresponds to small clusters, the value of $k_c$
is not evident. Clearly, the total number of clusters is significantly
smaller than the number of sites, and therefore $k_c\ll L^2$. Even a
tighter bound can be obtained by demanding the cutoff appears when the
cluster sizes reach a single site, i.e., $k_c\approx L^{2/q}B^{1/q}$. The
changing shape of the cutoff does not permit exact evaluation of its power
dependence on $L$ but it seems to increase slightly slower than $L^2$.

Both $P_k$ and $n_s$  describe the same cluster statistics from slightly different
points of view. Assuming that they both are power-laws, it should be possible
to relate them.  The cluster size $s$ for a specific fixed index $k$  fluctuates,
but it is possible to treat the mean of $s$ as a function of $k$. Similarly,
for a specific $s$ we can define the mean value of $k$. Both of these $s$--$k$
relations are expected to be the same power-laws. As long as the fluctuations
of $s$ for fixed $k$, or, alternatively, the fluctuations of $k$ for fixed $s$
are small, we can proceed with our derivation by assuming an approximate
deterministic relation $s(k)$. (For power-law distributions such a relation
can be used even when the fluctuations are large: one gets correct relations
between the exponents.) We will extract this relation from the mass of the
$k$th cluster $L^2P_k=L^2Bk^{-q}=s$. If there are $N_s$ clusters of size $s$,
the change of an $s$ cluster's mass by a unit advances its cluster index $k$ by $N_s$.
Therefore, we expect ${\frac{dk}{ds}=-N_{s}}$, or
\begin{equation}\label{eq:qtau_relate}
\frac{dk}{d\left(L^2Bk^{-q}\right)}=
-L^2As^{-\tau}=-L^2A(L^2Bk^{-q})^{-\tau}\ ,
\end{equation}
which relates the exponents $q$ and $\tau$ by
\begin{equation}\label{eq:qtau}
q=\frac{1}{\tau-1}\ .
\end{equation}

The slope of the graphs in Fig.~\ref{fig:Pk_graph} indicates that $q\approx 1.20$,
and this corresponds due to Eq.~\eqref{eq:qtau} to $\tau\approx1.83$.
The latter value is the same as the (approximate) measured value of
$\tau$ in the graphs of Fig.~\ref{fig:n_s_graph}. It is interesting
to note that that our heuristic estimate $\tau_{\rm th}=2$ corresponds
to the exponent $q=q_{\rm th}=1$. By examining the dependence of the
effective exponent $q$ on $1/\ln L$ we note that 1 is the likely
asymptotic value of $q$ for $L\to\infty$, although large error bars prevent
exact extrapolation. Further examination of the coefficients
in Eq.~\eqref{eq:qtau_relate} shows that the product of the prefactors
$AB^{1-\tau}\sim L^{2(\tau-2)}$. Since neither $A$ nor $B$ show significant
$L$ dependence, the anticipated asymptotic value $\tau=\tau_{\rm th}=2$
seems to be consistent with our results.

\section{\label{sec:conclusions}Conclusions and Discussion}
The study of RWs in $d=2$ is a very old and well explored subject.
We concentrated on percolation aspects of sites {\em not} visited
by the random walk on a periodic lattice both because this is the
lower critical dimension of a slightly more conventional percolation
problems of vacant sites in $d\ge3$, and because this problem exhibits
features that are absent in ``typical" lower critical dimension
problems. As far as it was possible, we used the tools of percolation
theory to analyze the problem, although certain features were very
different from regular percolation, and even different from
the behavior of Bernoulli percolation at its lower critical dimension.

Our paper was motivated by $d=2$ being the lower critical dimension of a
practically more important problem of gel degradation problem in $d=3$.
However, cross-linked membranes are ubiquitous in biology and medicine,
and their degradation, artificial or natural, plays an important role
warranting the consideration of degradation by an enzyme. Two-dimensional
processes with RW-controlled connectivity are also related to certain
models of photosynthetic bacteria colonies or photosynthetic membranes
in which the energy of light is transmitted to neighboring areas via
excitons thus creating  correlations (see Ref.~\cite{Rivoyre10} and references
therein). However, such correlations are short-lived and the problem
crosses over to Bernoulli percolation behavior~\cite{Auriac21}.

We have demonstrated that contrary to older results, the cluster
interiors are not fractal, although their boundaries are. The only
macroscopic scale of the 2DVSP problem is the lattice size $L$ and at
shorter distances the behavior is scale-free and can be described
using power-laws. The 2DVSP problem converges very slowly to the
``large $L$ limit." Our results indicate that the approach to
asymptotic behavior is as slow as the decay to zero of $1/\ln L$.
Consequently, all measurements even at large $L$ correspond to
intermediate effective behavior. We suggested a heuristic argument
setting the values of the exponents $\tau$ and $q$ describing the
cluster size distribution. Measured values of the exponents are
distinct but seem to move towards the theoretical values with
increasing $L$.

For $u\ll1$ the problem of cluster size distribution can be described
on an infinite lattice without the need to introduce periodic boundary
conditions. For $u\gg1$ the spanning clusters are virtually nonexistent
and one needs to understand clusters created by ${\cal N}_{\rm cr}\gg1$
almost independent pieces of the RW. Most of our measurements were performed
for $u=O(1)$, where the situation is most diverse, and might be different
from the behavior at very small or very large $u$s. Our paper does not
resolve this problem. Moreover, the approach to the limit of $p\to1$ is
much slower for larger $u$ values, and therefore various parts of the curves,
such as shown in Figs.~\ref{fig:Pi_by_u} to \ref{fig:Stilde_by_u}, converge at
different rates to their asymptotic values. Therefore the shape of the
curves may keep changing with increasing $L$.

The graphs in Fig.~\ref{fig:Pk_graph} have a rather distinct behavior for
$k\lesssim 4$ as opposed to larger $k$s. Such a separation into
``large" clusters (small $k$), and "small" clusters (larger $k$) provides
possible clues into the $u$ dependence of the features. By examining
similar graphs for  different values of $u$ we observed that increasing
$u$ decreases the sizes  of the ``large" clusters, but increases the sizes
of the clusters with larger $k$ towards the larger cluster sizes.
The $u$ dependence of all the properties requires a more systematic study.

Our theoretical arguments did not go beyond an approximate ``heuristic"
approach. However, a RW is a rather well understood object, and it is
conceivable that more accurate predictions can be made analytically.

\vspace{0.7 cm}

\begin{acknowledgments}
Y.K.~thanks M. Kardar for stimulating discussions. This work was supported
by the Israel Science Foundation Grant No.~453/17.
\end{acknowledgments}

\end{document}